\documentclass[10pt, conference]{IEEEtran}

\usepackage{xcolor}
\usepackage{xurl}
\usepackage{hyperref}
\usepackage{cite}
\usepackage[pdftex]{graphicx}
\usepackage{amsmath}
\usepackage{algorithmic}
\usepackage{array}
\usepackage[caption=false,font=normalsize,labelfont=sf,textfont=sf]{subfig}
\usepackage{stfloats}
\usepackage{url}
\usepackage{bbding}
\usepackage{pifont}
\usepackage{wasysym}
\usepackage{amssymb}
\usepackage{longtable,tabu}
\usepackage{supertabular,multicol}
\usepackage{comment}
\usepackage{multirow}
\usepackage{orcidlink}
\usepackage{booktabs,makecell}
\usepackage{tikz}
\def\BibTeX{{\rm B\kern-.05em{\sc i\kern-.025em b}\kern-.08em
    T\kern-.1667em\lower.7ex\hbox{E}\kern-.125emX}}
  
\newcommand{\linebreakand}{%
  \end{@IEEEauthorhalign}
  \hfill\mbox{}\par
  \mbox{}\hfill\begin{@IEEEauthorhalign}
}

\newcount\n
\def\tablebody{}
\makeatletter
\loop\ifnum\n<100
        \advance\n by1
        \protected@edef\tablebody{\tablebody
                \textbf{\number\n.}& shortText
                \tabularnewline
        }
\repeat

\makeatletter
\let\mcnewpage=\newpage
\newcommand{\TrickSupertabularIntoMulticols}{%
  \renewcommand\newpage{%
    \if@firstcolumn
      \hrule width\linewidth height0pt
      \columnbreak
    \else
      \mcnewpage
    \fi
  }%
}
\makeatother

\newif\ifdraft
\ifdraft
\newcommand{\note}[1]{ {\textcolor{orange} { **Note: #1 }}}

\newcommand{\jhanote}[1]{ {\textcolor{red} { ***Shantenu: #1 }}}
\newcommand{\alnote}[1]{ {\textcolor{green} { ***Andre: #1 }}}
\newcommand{\nsnote}[1]{ {\textcolor{blue} { ***Nishant: #1 }}}
\newcommand{\pnote}[1]{ {\textcolor{blue} { ***Pradeep: #1 }}}
\else
\newcommand{\note}[1]{}
\newcommand{\alnote}[1]{}
\newcommand{\nsnote}[1]{}
\newcommand{\jhanote}[1]{}
\newcommand{\crnote}[1]{}
\newcommand{\jknote}[1]{}
\newcommand{\pnote}[1]{}
\fi
\newcommand{\ignore}[1]{}

\newcommand{\upp}{\vspace*{-0.5em}}

\begin{document}

\bstctlcite{IEEEexample:BSTcontrol}

\title{A Conceptual Architecture for a Quantum-HPC Middleware}

\author{ \IEEEauthorblockN{Nishant Saurabh\IEEEauthorrefmark{1}\orcidlink{0000-0002-1926-4693},
Shantenu Jha\IEEEauthorrefmark{2}\IEEEauthorrefmark{3}\orcidlink{0000-0002-5040-026X}, 
Andre Luckow\IEEEauthorrefmark{4}\IEEEauthorrefmark{5}\orcidlink{0000-0002-1225-4062}\\}
\IEEEauthorblockA{\small\IEEEauthorrefmark{1}Utrecht University, NL\\
                  \IEEEauthorrefmark{2}Rutgers University, NJ, US\\
                  \IEEEauthorrefmark{3}Brookhaven National Lab, NY, US\\
                  \IEEEauthorrefmark{4}Ludwig Maximilian University Munich, Germany\\
                  \IEEEauthorrefmark{5}BMW Group, Munich, Germany\\
                  \IEEEauthorrefmark{1}n.saurabh@uu.nl,
                    \IEEEauthorrefmark{2}shantenu.jha@rutgers.edu,
                    \IEEEauthorrefmark{3}shantenu@bnl.gov,
                    \IEEEauthorrefmark{4}andre.luckow@ifi.lmu.de,
                    \IEEEauthorrefmark{5}andre.luckow@bmwgroup.com
                  \upp}\upp\upp\upp}
\maketitle
\begin{abstract}
  Quantum computing promises potential for science and industry by solving certain computationally complex problems faster than classical computers. Quantum computing systems evolved from monolithic systems towards modular architectures comprising multiple quantum processing units (QPUs) coupled to classical computing nodes (HPC). With the increasing scale, middleware systems that facilitate the efficient coupling of quantum-classical computing are becoming critical. Through an in-depth analysis of quantum applications, integration patterns and systems, we identified a gap in understanding Quantum-HPC middleware systems. We present a conceptual middleware to facilitate reasoning about quantum-classical integration and serve as the basis for a future middleware system. An essential contribution of this paper lies in leveraging well-established high-performance computing abstractions for managing workloads, tasks, and resources to integrate quantum computing into HPC systems seamlessly.

\end{abstract}


\section{Introduction}

Quantum computing promises to accelerate certain complex computations for science and industry applications. Quantum algorithms utilize quantum information sciences and promise speedups by requiring fewer steps than their classical counterparts. Many applications amenable to quantum computing traditionally utilize high-performance systems, simulations, machine learning,
and optimization techniques~\cite{bayerstadler2021industry}.

Algorithms for both \emph{Noisy Intermediate Scale Quantum Computers (NISQ)} and \emph{Fault-tolerant Quantum Computers (FTQC)} require the coupling of quantum and classical systems. For example, variational algorithms~\cite{vqa} depend on classical optimization and quantum error correction codes require significant classical computation of the syndrome measurements. 

The increasing maturity and modularity of quantum hardware is giving rise to the
question of integrating quantum systems into HPC systems~\cite{9537178,10007772, quantum_centric_hpc_2022,RUEFENACHT2022}. In the future, quantum computers will tightly integrate quantum processing units (QPUs) with classical HPC resources. Some NISQ devices are available in HPC centers~\cite{olcf} and the cloud~\cite{braket}. While the focus is often on hardware- and network-level integration, software integration has received less attention. 

In this paper, we explore the requirements of quantum applications and the need and types of integrations with classical HPC computing. We discuss and analyze 17 quantum application scenarios across optimization, machine learning, and simulation domains. The qualitative findings underscore the importance of categorizing these scenarios into three distinct integration patterns:
\emph{HPC-for-Quantum}, \emph{Quantum-in-HPC}, and \emph{Quantum-about-HPC}. For each scenario, we investigate different characteristics, such as the coupling of quantum and classical tasks and the resulting requirements for Quantum-HPC middleware. Our examination of the current state-of-the-art reveals significant limitations in the currently fragmented quantum software and middleware landscape, particularly concerning the ability to manage the application complexity and heterogeneous resources at scale.

To address these challenges, we design a conceptual middleware system that facilitates seamless interaction between classical (HPC) and quantum resources. Our conceptual middleware enables quantum-classical integration by providing unified resource access and management, enabling applications to allocate resources and manage quantum and classical tasks flexibly. The middleware utilizes proven high-performance computing abstractions to manage workloads, tasks, and resources to enable the seamless integration of quantum computing into HPC systems.

Using a chemistry application workflow as an example, we demonstrate the middleware system's ability to support the three integration patterns. This framework lays the foundation for developing a robust middleware system capable of managing and optimizing quantum applications in a unified Quantum-HPC computing environment.

This paper is structured as follows: Section~\ref{sec:applications} presents an overview of quantum applications and algorithms. We identify three integration patterns between classical (HPC) and quantum tasks in section~\ref{sec:quantum_hpc_integration}. We investigate how current quantum software and Quantum-HPC middleware systems address the challenges and needs of the identified integration patterns in section~\ref{sec:related}. We propose a conceptual middleware for  Quantum-HPC workflows in section~\ref{sec:quantum_hpc_middleware}. Section~\ref{sec:conclusion} concludes with a discussion of the results and future work.

\section{Applications and Algorithms}
\label{sec:applications}
In this section, we discuss the application scenarios and algorithms.

\subsection{Optimization}\label{subsec:optO}
Quantum algorithms, such as quantum annealing~\cite{das2005quantum} and quantum approximate optimization~\cite{farhi2014quantum}, are gaining traction in solving complex optimization problems.

\subsubsection{Use Cases}\label{ssubsec:optU} Optimization use cases include solving combinatorial optimization problems~\cite{farhi2014quantum} in scheduling~\cite{choi2020quantum}, logistics~\cite{awasthi2023quantum}, and transportation application domains~\cite{bayerstadler2021industry}. Quantum optimization also finds usage in quantum chemistry and drug discovery processes~\cite{cao2018potential,xiang2004fully}. Further, quantum computing can help optimize renewable energy system operations~\cite{li2019performance}, robotics (e.\,g., route optimization~\cite{gao2020advanced}),  and machine learning (e.\,g., optimizing trainable parameters for image classification and NLP~\cite{Schuld_2014}).

\subsubsection{Algorithms} such as quantum annealing (QA)~\cite{das2005quantum} map the optimization problem~\cite{bian2016mapping, pastorello2019quantum} onto an Ising model to explore the solution space. In contrast, Quantum Adiabatic Evolution (QAA)~\cite{schiffer2022adiabatic} encodes the optimization problem~\cite{perdomo2008construction} into the ground state of a known Hamiltonian. Further, it adiabatically evolves the system through a path of Hamiltonians to explore the solution space. Algorithms, such as the quantum approximate optimization algorithm (QAOA)~\cite{farhi2014quantum}, including enhancements like warm-starting~\cite{Egger_2021} and recursive QAOA (RQAOA)~\cite{PhysRevLett.125.260505}, use a combination of quantum and classical resources to find approximate solutions to combinatorial optimization problems. FTQC algorithms, such as Grover algorithm~\cite{grover_1996}, can also be used for optimization~\cite{gilliam2020optimizing}.

\subsection{Machine Learning}\label{subsec:optM}

Quantum computing can improve different parts of machine learning applications, e.\,g., linear algebra routines and generative machine learning methods. 

\subsubsection{Use Cases} Quantum machine learning (QML) use cases are often categorized based on the type of input data, i.\,e., classical, quantum, and the type of algorithm, i.\,e., supervised, unsupervised, and generative.  A typical QML use case with quantum data involves using quantum data for learning phases in many-body physics simulations~\cite{https://doi.org/10.48550/arxiv.2204.04198}.

QML applications using classical data~\cite{Schuld2021} include unsupervised clustering (e.\,g., the Sloan digital sky survey data, X-ray and weather data~\cite{weinstein2013analyzing}). Generative use cases of QML involve creating scientific simulation datasets (e.\,g., Monte Carlo events for particle physics process simulations~\cite{BravoPrieto2022stylebasedquantum, Rehm:2824092}) and chemical synthesis for molecular simulations~\cite{jain2022hybrid}. QML also finds usage in accelerating many-body Hamiltonian simulations~\cite{dqc}), enhancing quantum many-body simulations~\cite{liang20222,cai2018approximating,10.5555/3571885.3571948}, control in quantum routines~\cite{https://doi.org/10.48550/arxiv.2208.02645}, and optimizing quantum compilation~\cite{moro2021quantum}.

\subsubsection{Algorithms} FTQC algorithms, such as HHL~\cite{hhl} and QPE~\cite{qpe}, can be used for linear algebra subroutines (e.\,g., eigenvalues estimation, matrix inversion), distance computation between two quantum states~\cite{nielsen_chuang_2010} and finding closest neighbors~\cite{Schuld2021}. Unsupervised FTQC approaches, such as Q-means~\cite{q-means}, can identify data clusters and support the nearest centroid classification~\cite{quantum_inner_product}. 

There also exist variational algorithms for QML, e.\,g., variational quantum linear solvers (VQLS)~\cite{vqls}, differential quantum circuits (DQC)~\cite{dqc}, and quantum neural networks (QNN)~\cite{abbas2021power}. While VQLS and DQC are utilized for linear algebra problems, QNNs are used similarly to classical neural networks for classification problems, e.\,g., recognizing quantum states~\cite{qnn}. Quantum kernel methods~\cite{schuld_kernel_methods} learn a kernel function that maps data into a higher-dimensional Hilbert space using a variational circuit.

Generative QML techniques, such as Quantum Circuit Born Machines (QCBM)~\cite{e20080583, qcbm_quantum} and Quantum GANs (QGAN)~\cite{PhysRevA.98.012324}, demonstrated comparable training performance to classical models, requiring fewer parameters~\cite{riofrio2023performance}. They can generate data samples (e.\,g., input data for Monte Carlo simulations or generating new molecular states in quantum chemistry). QML is also applicable for surrogate modeling (e.\,g., FermiNet~\cite{pinn}).

\subsection{Simulation}\label{subsec:optS}
Quantum computers promise an exponential advantage in simulating quantum mechanical systems~\cite{feynman1982simulating}. For classic numerical simulations and modeling complex systems in science and engineering, promising algorithms, e.\,g., HHL~\cite{hhl}, exist.

\subsubsection{Use Cases} Application domains for simulation include material science, 
viz., the design of new materials, optimization of materials properties, and predicting material behavior. Quantum simulation methods can be used to
identify and design new compounds with desired medicinal properties,
predicting the potential side effects of drugs~\cite{xiang2004fully}.

Quantum computers can also be used for classical numerical
simulations~\cite{skamarock2001prototypes}, e.\,g., to study climate models~\cite{singh2021quantum}, to perform complex aerospace simulations~\cite{oz2022solving}, such 
as 3D computational fluid dynamics (CFD)~\cite{steijl2018parallel, joczik2022cost}, and to provide airflow predictions and estimate aircraft wing turbulence.

\subsubsection{Algorithms} 
Hamiltonian simulations~\cite{doi:10.1126/science.273.5278.1073} is the most well-known FTQC algorithm for simulating quantum-mechanical systems~\cite{doi:10.1021/acs.chemrev.8b00803,doi:10.1021/acs.chemrev.9b00829}. Hamiltonian simulations utilize further quantum subroutines, e.\,g., quantum phase estimation~\cite{qpe} for computing eigenvalues. There also exist NISQ approaches, e.\,g., the variational quantum eigensolver (VQE)~\cite{vqe}, and Quantum Monte Carlo (QMC)~\cite{Montanaro_2015,google_qmc}. While VQE utilizes a parameterized state with a classical optimizer to estimate the ground state of a Hamiltonian, QMC estimates the property of a quantum system using classical Monte Carlo methods and assesses the overlap between two quantum states on QPUs.

In the case of PDE-based numerical simulations, HHL~\cite{hhl} can be used to solve linear systems of equations, and QPE~\cite{qpe} for estimating the eigenvalues of a matrix; NISQ algorithms, such as DQC~\cite{dqc} and VQLS~\cite{vqls} also exist.

\subsection{Discussion}

\begin{table}[t]
\centering
\caption{{\bf Quantum Applications and Algorithms:} Common NISQ and FTQC algorithmic approaches proposed for different applications.}
\begin{tabular}{|@{ }c@{ }|@{ }c@{ }|@{ }c@{ }|@{ }c@{ }|@{ }c@{ }|@{ }c@{ }|@{ }c@{ }|}
\hline
\diaghead{Algo.App. class}{\bf Algo.}{\bf App class} & \emph{\bf Optimization} & \emph{\bf Machine} & \emph{\bf Quantum} & \emph{\bf Classical}\\
\emph &  & \emph{\bf learning} & \emph{\bf simulation} & \emph{\bf simulation}\\
\hline
\emph{Fault-tolerant} & Grover~\cite{grover_1996} & HHL~\cite{hhl}, Dist- & Hamiltonian & QPE~\cite{qpe}\\
\emph{algorithm} &  &ance estimation & simulation & HHL~\cite{hhl} \\
\hline
\emph{NISQ} & QAOA~\cite{farhi2014quantum} & QNN,QGAN & VQE~\cite{vqe} & VQLS~\cite{vqls}\\
\emph{algorithm} &  & QCBM &  & DQC~\cite{dqc}\\
\hline
\end{tabular}
\label{tab:quantum_computing_for_science_and_industry}
\end{table}

Quantum algorithms can benefit nearly all HPC applications. Many applications will likely emerge as incrementally quantum-enabled hardware and algorithms mature.  While the initial focus is on optimizing essential quantum algorithms, increasingly, the integration of these quantum-enhanced components in end-to-end application workflows needs to be considered. 

Table~\ref{tab:quantum_computing_for_science_and_industry} summarizes the discussed quantum algorithms in optimization, machine learning, and simulation. FTQC algorithms require many logical qubits, while the current capabilities of existing QPUs limit NISQ algorithms. These typically possess only a few noisy qubits with limited coherence times.

All quantum algorithms will be hybrid~\cite{callison2022hybrid}, i.\,e., a significant computation part is done on classical resources. In particular, for NISQ, only critical kernels that provide decisive quantum advantages will be run on a QPU. We expect these quantum kernels to be highly algorithm- and hardware-dependent (e.\,g., qubit modality, simulator, interconnect). 

Increasingly, algorithms from all three domains are used together, e.\,g. simulation output data is used for machine learning or as input for prescriptive optimization. Further, ML-generated data frequently serves as input to simulations.

\section{Integration Patterns}
\label{sec:quantum_hpc_integration}

While quantum computers can encode any function that a classical computer can, running complete workflows on quantum computers will soon not be feasible due to the high depth and qubit count required. Thus, quantum applications will need to contend with hybrid resources. Only a minimal kernel, providing a quantum advantage, will often be executed on a standalone QPU. These kernels will be augmented with significant classical components (both for NISQ and FTQC). Hence, a better understanding of the interaction between classical and quantum components requires investigating their integration patterns and analyzing the types of coupling and the application structure.

We identify three types of integration between classical and quantum
tasks: \emph{HPC-for-Quantum}, \emph{Quantum-in-HPC}, and \emph{Quantum-about-HPC}. We investigate two main characteristics: the coupling~\cite{MCCASKEY2018245} and the application structure.  

\begin{table*}[t]
  \centering
  \caption{{\bf HPC-for-Quantum Application Scenarios} {\normalfont categorized by coupling and application structure, focusing on the different classical and quantum tasks. Significant classical computation is required to support a quantum computer.}}
  \begin{tabular}{|@{ }c@{ }|@{ }c@{ }|@{ }c@{ }|@{ }c@{ }|@{ }c@{ }|@{ }c@{ }|@{ }c@{ }|}
  \hline
  \emph{\bf Scenario} & \emph{\bf Description} & \emph{\bf Coupling} & \emph{\bf Category} & \emph{\bf Application} & \emph{\bf Classical Task} & \emph{\bf Quantum Task}\\
  \emph &  &  &  & \emph{\bf structure} &  &\\
  \hline
  \emph{Quantum} & Controlling adiabatic/ & tight & NISQ &Accelerators &Bayesian Optimization/ & All QCs\\ 
  \emph{control} & diabatic schedule &  &  &  & Reinforcement Learning &\\
  \hline
  \emph{Error-mitigation} & Embedded into algorithm     & tight & NISQ/ &Accelerators & Surface codes  & All QCs\\
  \emph{\& correction~\cite{PhysRevLett.108.180501}} &  with repeated measurements  &  & FTQC & & with significant  &  \\
  \emph &  \& application corrections &  &  & & classical processing &  \\
  \hline
  \emph{Dynamic circuits} & Circuits that are conditioned  & tight & NISQ &Accelerators & classical processing  & All QCs\\
  \emph & on the input of &  & &  & of auxiliary qubit with  & \\
  \emph & real-time classical components  &  & &  & feedback into circuit & \\
  \hline
  \emph{Circuit} & Decomposing large quantum  & tight & NISQ & Task & Circuit decomposition  & All QCs\\
  \emph{knitting~\cite{PhysRevLett.125.150504}} & circuits into smaller circuits for  &  & & Parallelism & \& result collection & \\
  \emph & distribution across QPUs &  & &  & & \\
  \hline
  \emph{Classic} & Using HPC resources, & - &Classical &Ensemble, Task & Statevector, tensor  & n/a\\
  \emph{simulation~\cite{cuquantum}} & methods to simulate  &  &  &Parallelism,  & networks, density   & \\
  \emph & quantum computers &  &  & Accelerators & matrix simulation &\\
  \hline
  \end{tabular}
  \label{tab:quantum-for-hpc}
  \end{table*}

\emph{Coupling:} Coupling describes the time sensitivity of the interaction between components. The coupling of tasks can occur tightly within the coherence time of the QPU, i.\,e., the time that a QPU can maintain its state, in near time, for example, to post-process measurements (i.\,e., medium), and in end-to-end application workflows, (i.\,e., loose)~\cite{MCCASKEY2018245}. 

We refer to \emph{tight-coupling} if tasks need to interact within strict time-sensitive bounds, e.\,g., within the QPU coherence window.  Examples of tight coupling are quantum error mitigation, error correction, and algorithms that utilize mid-circuit measurements. 

\emph{Medium-coupled} scenarios comprise tasks that require frequent, time-sensitive interaction, but coupling between QPU at coherence time is not needed. The coupling of quantum and classical tasks happens outside the coherence time of the quantum computer. Examples are variational algorithms, such as VQE and QAOA, which process the measurements after each execution of the circuit.

In \emph{loosely-coupled} scenarios, less frequent interaction is needed, e.\,g., the results are processed together after the parallel job. Loose coupling refers to a coupling on the workflow layer that integrates seamlessly across quantum, classical, and hybrid components of tasks and their dependencies.

The \emph{application structure} describes how the application exploits various types of parallelism, e.\,g., ensemble, task parallelism, data parallelism, and accelerators.  Quantum programs can expose different types of parallelism, both in the classical and quantum parts. For example, quantum algorithms typically involve repeated sampling from circuits, i.\,e., the circuit must be repeatedly executed and, thus, are amenable to parallelism.   Circuit knitting allows the partitioning and parallel execution of circuit parts on multiple QPUs.

\subsection{HPC-for-Quantum}
The HPC-for-Quantum integration pattern describes the usage of classic compute and HPC techniques on the low-level system layer to accommodate I/O, dynamic circuits, error mitigation, and other techniques that enable the most effective utilization of the QPU. The layer primarily concerns low-level circuit developers that ensure the execution of quantum circuits on the QPU (Ref.~\cite{quantum_centric_hpc_2022}). HPC and quantum tasks are tightly connected and interact in real-time. Approaches are mostly application-agnostic, e.\,g., error correction, compilation, and parallelism. HPC techniques, e.\,g., parallelization and acceleration using GPUs and FPGAs and high-performance networking, can provide significant advantages to the quantum kernel and application.

Table~\ref{tab:quantum-for-hpc} summarizes scenarios for the HPC-for-Quantum integration type. HPC support is crucial to support quantum control, error mitigation, and error corrections, as well as dynamic circuits that require tight coupling of classic and quantum tasks.

HPC technologies are increasingly essential for quantum control systems and enable the optimal manipulation of the qubits through physical operations~\cite{ella2023quantumclassical}. For example, determining the optimal timing and method of sending pulses to control the qubits is computationally expensive. Scalable approaches have been proposed, e.\,g., Quandary~\cite{https://doi.org/10.48550/arxiv.2110.10310} uses MPI to distribute necessary computations.    

Further, many quantum error mitigation and correction aspects are computationally intensive and require tight integration~\cite{ella2023quantumclassical}. For example, the error mitigation of quantum circuits can be performed by running multiple noisy experiments so that errors cancel out~\cite{quantum_centric_hpc_2022}.

Dynamic circuits involve both the evolution of the quantum state and mid-circuit measurements. The measurements must simultaneously be processed classically  (i.\,e., within coherence time) and are used to steer further quantum processing, e.\,g., by branching or setting variables. Corcoles et\,al.~\cite{PhysRevLett.127.100501} demonstrated algorithm improvements, e.\,g., in QPE.
 
Classic simulations of quantum systems are also an essential building block and critical for evaluating quantum algorithms and hardware.  HPC and AI techniques provide the necessary scale to simulators, e.\,g., by using parallel and GPU-accelerated simulators (e.\,g., cuQuantum~\cite{cuquantum} for multi GPU and node state vector simulation) and task-parallel tensor network simulations~\cite{Vincent_2022}.

\ignore{
  Several HPC techniques play a crucial role in supporting the above-mentioned
  quantum application scenarios, specifically, the utilization of compilers,
  parallelism (including accelerators), and high-end networks.}

\subsection{Quantum-in-HPC}

\begin{table*}[t]
\centering
\caption{{\bf Quantum-in-HPC Application Scenarios:} {\normalfont FTQC and NISQ algorithms require coupling with classical tasks for pre-, post- and optimization tasks.}}
\begin{tabular}{|@{ }c@{ }|@{ }c@{ }|@{ }c@{ }|@{ }c@{ }|@{ }c@{ }|@{ }c@{ }|@{ }c@{ }|}
\hline
\emph{\bf Scenario} & \emph{\bf Description} & \emph{\bf Coupling} & \emph{\bf Category} & \emph{\bf Application} & \emph{\bf Classical Task} & \emph{\bf Quantum Task}\\
\emph &  &  &  & \emph{\bf structure} &  &\\
\hline
\emph{Hamiltonian} & Time evolution of  & - & FTQC &- & - & Hamiltonian\\ 
\emph{Simulation~\cite{doi:10.1126/science.273.5278.1073}} & Schrödinger's equation. &  &  &  &  & simulation\\
\hline
\emph{Quantum Phase} & Subroutine to extract & medium & FTQC & Task & Different variants of QPE  & QPE\\
\emph{Estimation (QPE)~\cite{qpe}} &  eigenvalue and eigenstates  &  & /FTQC & Parallelism & with different levels of classical   & circuit  \\
\emph & from a Hamiltonian.&  &  & & processing interweaved~\cite{PhysRevLett.127.100501} &  \\
\hline
\emph{Imaginary Time} & Variational algorithm  & medium & NISQ & Task & linear equations  & Trial\\
\emph {Evolution~\cite{qite}} & utilizing trial quantum state  &  & & Accelerator & solving  & state\\
\emph & classic linear solver  &  & &  &  & \\
\hline
\emph{Variational} & Quantum kernels & medium & NISQ & Task & Optimization loop, warm-  & Parameterized circuit \\
\emph{Algorithms~\cite{vqa}} &  with classical optimizers &  & & Parallelism & starting, pre-/post-processing, & (ground state estimation,\\
\emph & (VQE, QAOA, QNN) &  & &  & RQAOA elimination~\cite{PhysRevLett.125.260505} & QAOA circuit)\\
\hline
\emph{Generative} & QCMB, QGANs (quantum  & medium & NISQ & Task & Optimizer,  & Copula circuit,\\
\emph{AI~\cite{qcbm_quantum}} & generator with classic &  &  & Parallelism,  & Discriminator  & strongly entangled\\
\emph & or quantum discriminator) &  &  & Accelerators & module & circuit\\
\hline
\end{tabular}
\label{tab:quantum-in-hpc}
\end{table*}

\emph{Quantum-in-HPC}  is an integration pattern where a quantum component is medium-coupled with a classical HPC component. In contrast to HPC-for-Quantum scenarios, e.\,g., dynamic circuits, the interweaving of classical and quantum computation does not occur in real-time during the coherence of the QPU, but after each measurement cycle.

Table~\ref{tab:quantum-in-hpc} summarizes different scenarios. Typically, these scenarios involve an ensemble of quantum tasks for repeated measurements.  The orchestration of the application typically resides on the classical HPC system, with the QPU only providing acceleration for the different types of parameterized quantum circuits (PQC).

Examples where this is already the case are variational quantum algorithms (VQAs)~\cite{vqa}, e.\,g., VQE for estimating the ground state energy of a molecular system,  QAOA for solving combinatorial optimization problems, VQLS~\cite{vqls} for solving linear equations, or quantum GANs for generative AI~\cite{qcbm_quantum}. The amount of classical computing can vary significantly from a classical optimizer loop to comprehensively pre-computing states, e.\,g., using warm starting procedures (see Quantum-about-HPC integration type).

The middleware must manage different types of parallelisms. For example, ensembles, i.\,e., multiple independent tasks executed on a QPU, and more general task parallelism, where complex task dependencies must be handled. For example, VQAs exhibit more complex task parallelism, where each generation of quantum tasks depends on the results of the previous generation.

\subsection{Quantum-about-HPC}
\begin{table*}[t]
\centering
\caption{{\bf Quantum-about-HPC Application Scenarios} {\normalfont enable the integration of quantum application components into end-to-end workflows.}}
\begin{tabular}{|@{ }c@{ }|@{ }c@{ }|@{ }c@{ }|@{ }c@{ }|@{ }c@{ }|@{ }c@{ }|@{ }c@{ }|}
\hline
\emph{\bf Scenario} & \emph{\bf Description} & \emph{\bf Coupling} & \emph{\bf Category} & \emph{\bf Application} & \emph{\bf Classical Task} & \emph{\bf Quantum Task}\\
\emph &  &  &  & \emph{\bf structure} &  &\\
\hline
\emph{Classic} & Encode classical data  & loose & NISQ & Workflow & Data embedding for ML & Quantum-in-HPC \\ 
\emph{preprocessing} & into a quantum state,  &  &  &  & ~\cite{Schuld2021}, molecular Hamiltonian & application\\   
\emph & e.\,g., in QML and  &  &  &  &  and initial parameters using & component\\
\emph & Quantum Chemistry &  &  &  &  Hartree-Fock method~\cite{doi:10.1021/acs.chemrev.8b00803} & (Table~\ref{tab:quantum-in-hpc})\\
\hline
\emph{Classic} & Partial extraction of quantum computation & loose & NISQ & Workflow & Parallel processing of  & Same as\\
\emph{post-processing} &  results via QPU measurements. Classic &  & /FTQC &  & expectation values from  & above  \\
\emph & post-processing for reconstructing &  &  & & observable computed   &\\
\emph & state for further processing. &  &  & & on a QPU  &\\
\hline
\emph{Hyperparameter} & Select optimal parameters  & loose & NISQ & Ensemble & Parameter selection  & Same\\
\emph {optimization~\cite{9973678}} & for quantum kernel (e.\,g., cost function,   &  & &  & \& post-processing  & as above\\
\emph & learning rate, initialization)  &  & &  &  & \\
\hline
\emph{AI} & QML to learn complex states as input for & loose & NISQ & Workflow & Classic simulation loop,  & Same as \\
\emph{workflows} &   simulation and property prediction &  & &  & pre/post-processing & above\\
\emph & (chemistry~\cite{generative_molecule_design_2022} and optimization~\cite{https://doi.org/10.48550/arxiv.2101.06250}) &  & &  & optimization &\\
\hline
\emph{Warm} & Warm-starting of quantum  & loose & NISQ & Workflow & Heuristics (MILP, & Same as\\
\emph{starting} & algorithm with classical  &  &  &   &  CPLEX) to pre-compute & above\\
\emph & solution~\cite{Egger_2021} &  &  &  & initial parameters & \\
\hline
\emph{Hybrid Quantum} & Complex part (e.\,g., sign problem) & medium & NISQ & Task & Sample generation,  & Overlap esti-\\
\emph{Monte Carlo~\cite{google_qmc}} & on QPU (overlap between sample and &  &  & Parallelism  & time  & mation between\\
\emph & trial wave function), executing &  &  &  & evolution & trial and sample\\
\emph & other parts classically (time evolution) &  &  &  &  & wave function \\
\hline
\emph{Quantum-Quantum} & Coupling Hamiltonian simulation & tight & FTQC & Workflow & Optimizer,  & Hamiltonian\\
\emph{coupling} & and analysis of  &  &  &  & Discriminator  & simulation\\
\emph & static properties results &  &  & & module &\\
\hline
\end{tabular}
\label{tab:quantum-about-hpc}
\end{table*}

The \emph{Quantum-about-HPC} integration pattern describes scenarios where a quantum-enhanced kernel is integrated into an end-to-end quantum-classical workflow. In other words, the quantum component is added without modifying the HPC application -- unlike in the Quantum-in-HPC integration pattern. In this case, the quantum component is used as a black box but requires further input or output to be effective. The main application control loop generally resides in the classical system. On this level, the quantum and classical components are often looser coupled than with other integration types.

Table~\ref{tab:quantum-about-hpc} summarizes different application scenarios for the Quantum-about-HPC integration type. For example, pre- and post-processing tasks commonly need to be performed, e.\,g., for data encoding, loading and converting data, and pre-conditioning quantum algorithms. An example of pre-conditioning is warm-starting QAOA, for which a classic solution determines the initial parameters. In other cases, the quantum results are inputs for further classic or quantum processing. For instance, the output of quantum generative methods (e.\,g., QGAN, QCBMs) serves as input for further optimization and numerical simulations (e.\,g., in quantum chemistry and high-energy physics~\cite{Rehm:2824092}).

However, such workflows can exhibit more complex integration patterns, e.\,g., the in-situ processing of quantum data with quantum machine learning (ML), the training of ML surrogate models to mitigate the data readout bottleneck, and the coupling of generative Quantum ML with other techniques, e.\,g., simulation and optimization (both quantum and classical).

\subsection{Discussion}

\begin{figure}
    \centering
    \includegraphics[width=\linewidth]{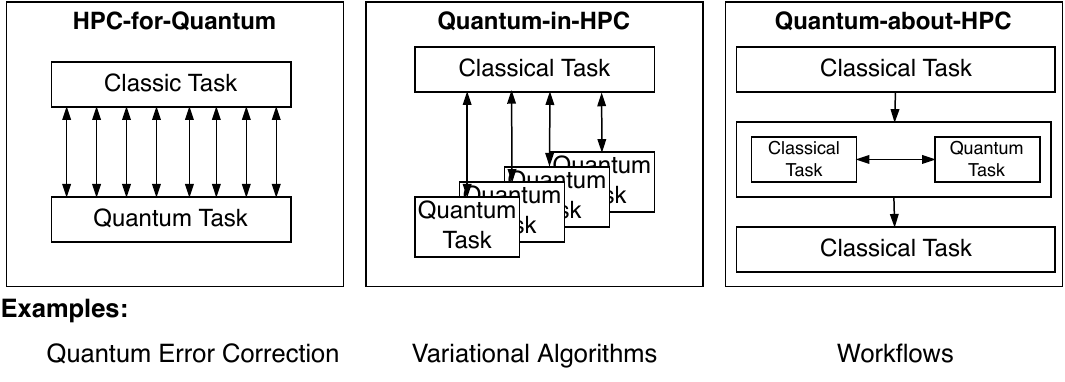}
    \caption{\textbf{Quantum-HPC Integration Patterns: }HPC-for-Quantum requires interactions within the coherence time of the QPU, Quantum-in-HPC utilizes a classical task to orchestrate short-running quantum tasks, Quantum-about-HPC connects composable tasks to workflows.}
    \label{fig:quantum-hpc-integration-patterns}  
\end{figure}

Figure~\ref{fig:quantum-hpc-integration-patterns} summarizes application patterns for the different Quantum-HPC integration types. The HPC-for-Quantum scenarios describe the tight integration of HPC and quantum resources, often in real-time, i.\,e., within the coherence time of the quantum system. It is characterized by frequent data exchange between the classical and quantum systems to process mid-circuit measurements, e.\,g., for error correction.

Quantum-in-HPC describes interactions where the QPU is integrated as an accelerator for specific task types, e.\,g., for the evaluation of a quantum state. The classical task is long-lived, maintaining the overall state and utilizing the QPU for specific quantum tasks. These are short-lived compared to the classical task. As described in Table~\ref{tab:quantum-in-hpc}, variational algorithms have been proposed for nearly all problem domains, e.\,g., for machine learning, optimization, and linear algebra.
Finally, Quantum-about-HPC workflows integrate heterogeneous tasks, i.\,e., quantum, classical or composable tasks, into end-to-end application scenarios. Workflows comprise distinct stages, e.\,g., data collection, pre-/post-processing, and simulation.

Managing quantum and classical resources can be difficult due to the varying and unpredictable resource demands, requiring a sophisticated approach to resource management.  For example, the QPU resource demands for variational circuits can vary significantly as using different optimizers, e.\,g., can result in a different number of circuit executions. Gradient-based optimizers require more executions of a quantum circuit to estimate the gradient using the parameter shift rule than non-gradient-based optimizers. Thus, a middleware system that can adaptively manage the resources is required. With scale, data and computational requirements will become even more demanding, exacerbating the need for careful resource management.

\section{State of the Art and Related Work}
\label{sec:related}

This section describes the current ecosystem of quantum software
frameworks, Quantum-HPC integration, and middleware systems. Mainly, we investigate 
how these systems address the challenges related to integration patterns, as described in Section~\ref{sec:quantum_hpc_integration}. 

\subsection{Quantum Software Libraries}
\label{sec:quantum-software}

Various quantum software frameworks emerged, e.\,g., Pennylane~\cite{bergholm2018pennylane}, Qiskit~\cite{cross2018ibm}, Cirq~\cite{hancockcirq}, Intel Quantum SDK~\cite{khalate2022llvmbased,Guerreschi_2020}, Quil~\cite{robert_s_smith_2020_3677541} and Quantum Brilliance SDK~\cite{nguyen2022software}. Here we summarize the key aspects and limitations of these quantum software frameworks. For a detailed survey, refer to Serrano et\,al.~\cite{10.1145/3548679}.

The existing frameworks support creating and executing quantum circuits on multiple quantum backends (e.\,g., simulators and real quantum devices) and enable interfacing with various hardware platforms (e.\,g., superconducting and ion trap platforms). Further, several high-level libraries that provide ready-to-use quantum-based algorithms for optimization, machine learning, and simulation have been developed. For QML applications, utilizing a gradient-based optimizer with quantum circuits is critical. For example, Pennylane supports differentiable quantum circuits by integrating with machine learning frameworks like PyTorch, Jax, and Tensorflow. 

Most frameworks also provide parallelization and accelerator support, e.\,g., for just-in-time compilation and accelerated GPU simulators. However, they are often limited to specific cloud platforms and must better interface with HPC resource managers. As workload and task management are deeply integrated into these frameworks, the degree of integration with HPC systems and, thus, the scale is limited.

\begin{comment}
\subsection{Hardware}

A quantum computer is a device that combines both classical and quantum components. The QPU (quantum processing unit) is the core component of a quantum computer and is used to perform quantum computing operations. There are several models of quantum computing, the most popular ones are gate-based and analog. A gate-based quantum computer uses standardized quantum gates, such as single-qubit and two-qubit gates to control quantum states. Analog models, on the other hand, use adiabatic/diabatic methods to simulate quantum systems. Examples of gate-based devices include IBM, Google, and Rigetti's superconducting systems, as well as ion trap systems from IonQ and Quantinuum. Analog neutral atom systems examples include QuEra and Pasqal. These systems currently offer an analog simulation of the bloqade Hamiltonian on a Rydberg qubit device, but in the future, a more general gate-based abstraction will be available for these systems.
\end{comment}

%
%

%

%

\subsection{Quantum-HPC Integration}

Current trends show a transition from remote cloud access for QPUs to a more tightly integrated HPC model, wherein the QPU is co-located alongside classical HPC computing resources~\cite{9537178, 10007772, quantum_centric_hpc_2022, RUEFENACHT2022}. 

Minimizing the latency and bandwidth is critical for HPC-for-Quantum and HPC-in-Quantum use cases. Ella et\,al.~\cite{ella2023quantumclassical} investigate low-level HPC-for-Quantum integration on pulse-level required for quantum control, error correction, and mitigation. They emphasize the need for HPC classical resources, including accelerators, to be co-located with the QPU.

Another vital concern, particularly for HPC-in-Quantum use cases, is resource management. While existing resource management systems, e.\,g., SLURM, can support QPUs (e.\,g., using SLURM's Generic Resource abstraction), but there are significant limitations, e.\,g., regarding support for different QPU types and the prioritization of QPU jobs. Further, application-level workloads and task management systems, such as Pilot-Jobs~\cite{6404423} and Hyperqueue~\cite{jakub_beranek_2023_7838764}, must be integrated with quantum software frameworks. Ruefenach et al.~\cite{RUEFENACHT2022} summarizes many of these challenges, e.\,g., ensuring the optimal utilization of the QPUs, while minimizing the time-to-solution and energy-to-solution, and propose a quantum resource manager.

\subsection{Quantum-HPC Middleware}

Integrating quantum and classical tasks and quantum-HPC middleware systems are subject to intense research. For example,  XACC~\cite{xacc_2020} introduces a quantum-classical programming model that allows tighter integration between both computing paradigms. CUDA Quantum~\cite{cuda_quantum} is a platform for integrating classical and quantum computing devices using a standard programming model similar to XACC. It offers optimized control and communication between different quantum processors and classical tasks. It integrates with the cuQuantum GPU library for accelerated simulations with scaling across distributed multi-node and multi-GPU systems. Further, support for selected QPUs (e.\,g, Rigetti) will be available. 

Increasingly, hybrid quantum-classical runtimes are integrated into existing quantum software frameworks. For example, Qiskit Runtime~\cite{johnson2022qiskit} and Braket Jobs~\cite{braket-jobs-2021} provide mechanisms to manage classical computing with quantum tasks more efficiently. These are limited to proprietary cloud environments.

Quantum workflows have become critical in addressing the need to integrate quantum components into end-to-end applications (Quantum-about-HPC integration type). Weder et\,al.~\cite{9302814, Weder2022} investigate workflow technology for orchestrating quantum applications. \emph{Tierkreis~\cite{https://doi.org/10.48550/arxiv.2211.02350}} focuses on task parallelism exhibited by hybrid quantum-classical applications and utilizes a dataflow-based programming model. Other commercial tools emerged, e.\,g., Orquestra~\cite{zapata2021orchestra} and Covalent~\cite{covalent2023}.

\section{Quantum-HPC Middleware: Toward a Conceptual Framework}
\label{sec:quantum_hpc_middleware}

In this section, we define the functional layers for a Quantum-HPC middleware,
identify challenges and design objectives for each layer, and describe a
conceptual middleware.

\subsection{Functional Layers}

\begin{figure}[t]
  \centering
  \includegraphics[width=0.5\textwidth]{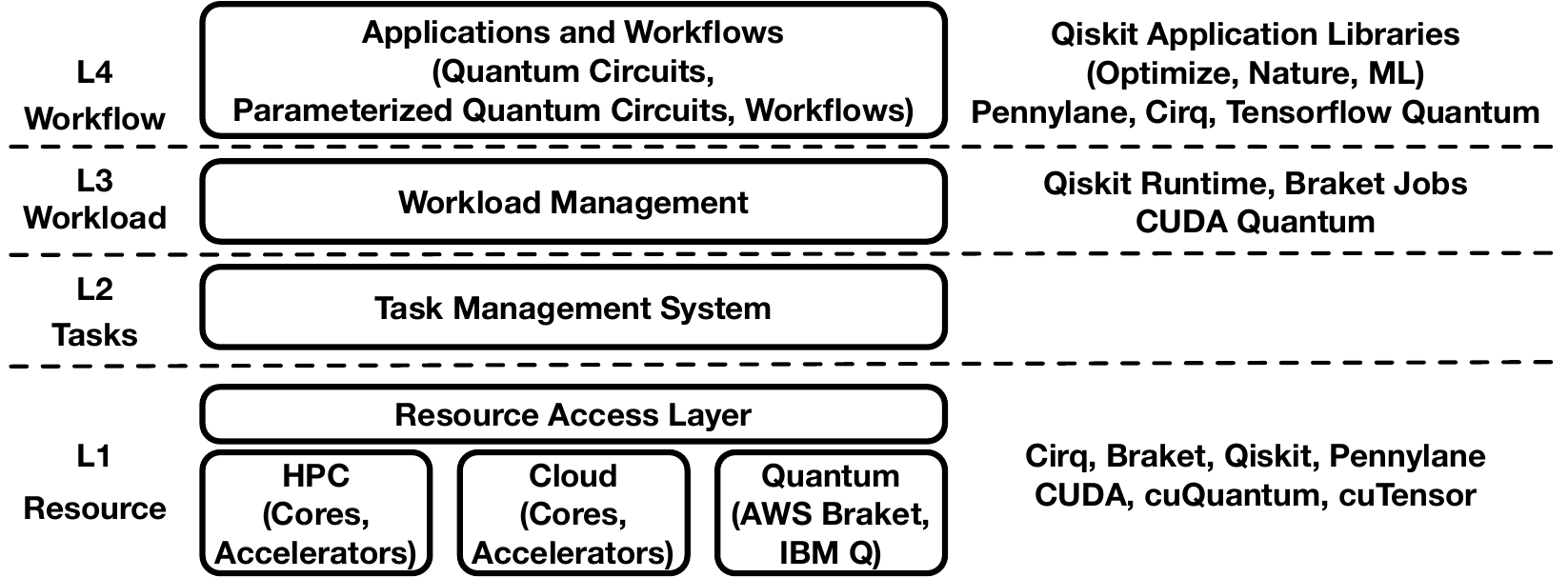}
  \caption{\textbf{Functional Levels for Quantum-HPC Middleware (adapted from~\cite{Turilli_2019}):} The middleware system can be partitioned into four layers. The workflow layer encapsulates the application semantics and logical dependencies between the different workflow stages. The workload layer translates the workflow into a set of tasks that can be executed, potentially concurrently. The task layer executes these tasks using the resource layer.} 
  \label{fig:functional-levels}
\end{figure}

We adopt the four-layered model for managing scientific workflows on HPC resources proposed by Turilli et\,al.~\cite{Turilli_2019}. Figure~\ref{fig:functional-levels} illustrates the functional layers: the resource (L1), task (L2), workload (L3), and workflow (L4) layers. Further, it shows different quantum software libraries alongside appropriate layers.

The highest layer is the workflow layer (L4) which encapsulates the application semantics and logical dependencies between the different workflow levels. It is the most abstract layer and is often exposed using a domain-specific language (DSL) to describe the workflow. The workflow manager translates the workload description into a workload, i.\,e., a set of tasks that can be executed, potentially concurrently.

The workload and task layers are the middle Layers (L2, L3). The workload layer (L3) selects the appropriate resources for the given workload. The task layer (L2) executes these tasks on the selected resource. For this purpose, it includes functionality to acquire respective HPC resources (e.\,g., a Pilot-Job~\cite{6404423}). For quantum computing, the co-allocation of quantum and classical resources is critical.

The resource layer (L1) is responsible for scheduling and assigning computational tasks to the various resources within the HPC system, such as nodes, processors, and QPUs. In particular, in the context of quantum computing, the heterogeneity at this layer is challenging. Advances in the intermediate representation, e.\,g., the Quantum Intermediate Representation (QIR)~\cite{qir_alliance} and Open Quantum Assembly Language (QASM)~\cite{OpenQASM}, and unified access APIs are critical to ensure uniform access to heterogeneous hardware. 

On the resource level (L1), the focus is on executing quantum tasks (i.\,e., quantum circuits) and related classical tasks on quantum/classical resources. The execution of quantum circuits involves compilation, error mitigation/correction, measurements, and other low-level optimization steps. Further, repeated measurements to obtain a representative sample of the quantum state and post-processing (e.\,g., to compute expectation values) are managed on this layer.

\subsection{Challenges and Design Objective}

This section identifies challenges and design objectives (O) at each functional layer that can lead to high-level Quantum-HPC middleware architecture.

\paragraph{L4 -- Workflow Layer ({\bf \textbf{O-1}})} 
Applications require integrating a diverse set of classical (e.\,g., classical AI and HPC tasks) and quantum tasks (e.\,g., simulated quantum tasks) in the end-to-end workflows. Achieving such objectives requires modular and composable architectural designs to enable re-use at different levels, e.\,g., the function, library, and system levels. Additionally, integrating diverse quantum software libraries and components is important~\cite{Turilli_2019}. Quantum workflows possess additional complexity, e.\,g., they require incorporating different types of QPUs (simulated, ion-traps, superconducting). Often, applications involve, e.\,g., simulated and different physical QPUs, and, thus, require particular adaptations while the application logic remains the same. The high-level workflow description is then converted into a workload comprising heterogeneous tasks that must be mapped to a complex infrastructure of nodes, CPUs, GPUs, and QPUs. In particular, the software ecosystem is highly fragmented in the current (early) stage, and standards for describing and executing quantum workflows are missing.

\paragraph{L3, L2 -- Workload and Task Layer ({\bf \textbf{O-2}})} 

Workloads consist of highly heterogeneous containing components and tasks implemented in different languages and frameworks as described previously. For instance, quantum machine learning requires the integration of quantum frameworks, like Pennylane, with machine learning frameworks, such as PyTorch and Jax. Application resource requirements can vary significantly with specific configurations (e.\,g., the QPU type or optimizer in a VQA). Often the same workload must be executed at different scales and on other resource types (e.\,g., classical simulators and QPU)s, leading to significantly different execution characteristics on the workload and task layer. For example, the execution time and results of the same quantum tasks can vary considerably with the QPU type (simulated vs. physical QPU of different modalities)~\cite{lubinski2023optimization}.  A challenge is to identify emerging workload patterns that need to be supported by the workload management system.

\paragraph{L1 -- Resource Layer ({\bf \textbf{O-3}})} 
The resource layer encapsulates the heterogeneous quantum and classical resources. Challenges arise concerning integrating quantum resources and supporting the tight coupling of quantum and classical tasks, e.\,g., for quantum error correction and dynamic circuits. Tight coupling requires the co-allocation of resources to ensure frequent and low-latency interactions between quantum and classical tasks. A tighter integration at the hardware and network level is desirable, particularly for HPC-for-Quantum scenarios. While in the traditional accelerator model, GPUs are dedicated to a single application, the scarcity of physical QPUs requires more careful consideration of resource allocation.

\subsection{Conceptual Middleware}

\begin{figure}[t]
  \centering
  \includegraphics[width=0.5\textwidth]{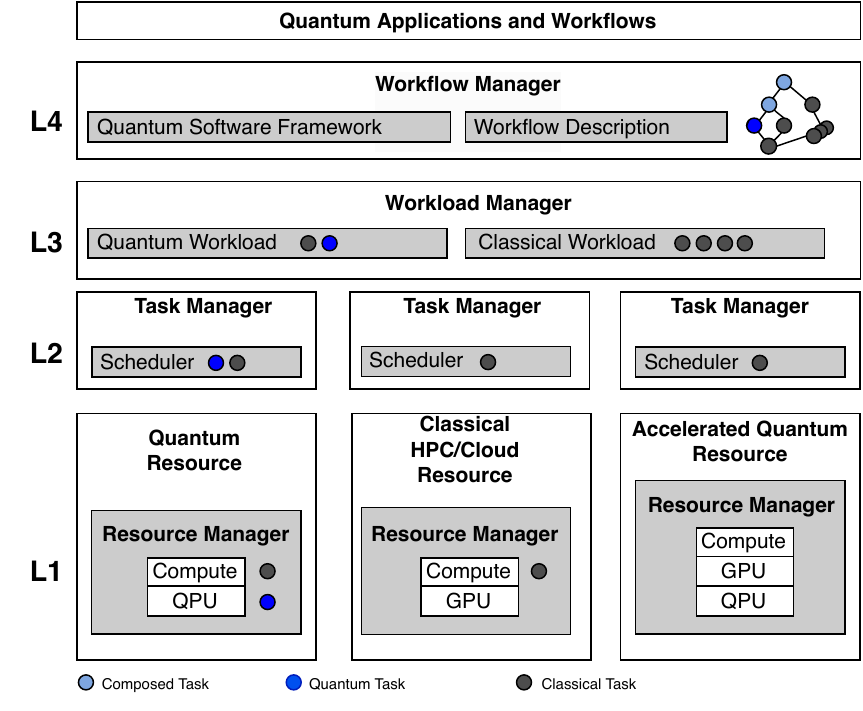}
  \caption{\textbf{Conceptual Quantum-HPC Middleware:} The Quantum-HPC middleware is composed of four layers: workflow, workload, task, and resource layers. By decoupling application, workload, task, and resource management concerns, the middleware enables the necessary scale for next-generation Quantum-HPC systems.}
  \label{fig:quantum-concept}
\end{figure}

Figure~\ref{fig:quantum-concept} illustrates the conceptual Quantum-HPC middleware with four layers, namely, workflow, workload, task, and resource layers, based on the functional levels identified in Figure~\ref{fig:functional-levels}.

\subsubsection{Workflow Layer} 

The workflow layer offers a high-level abstraction for quantum workflows, containing both quantum and classical components. It receives high-level descriptions of dependencies, input/output data, and computational tasks. The workflow manager coordinates workflow descriptions and prepares their execution.

We identify three task types: classical, quantum, and composite tasks. Classical tasks are self-contained classical computations, such as data loading, pre-processing, and post-processing. Quantum tasks are self-contained quantum circuits, executable on a QPU. Quantum workloads contain both classical and quantum tasks.

A quantum task refers to self-contained quantum computation, also called a circuit, that is executable on a QPU. A quantum task can be defined using a low-level language (e.\,g., OpenQASM) or a quantum software framework (e.\,g., Qiskit). Quantum tasks are typically coupled to different classical computational tasks, e.\,g., for post-processing, quantum error correction, and hybrid algorithms. 

Composite tasks consist of multiple sub-tasks, often integrating external software frameworks, which is crucial for quantum workflows. For example, a composite task can arise by integrating an external software framework, an essential requirement for quantum workflows. For example,  Qiskit provides several optimization, simulation, and machine learning application frameworks. Using composite tasks, a Qiskit QAOA implementation can be integrated into a workflow with further pre- and post-processing steps.

The workflow manager transforms the workflow into an executable state by resolving dependencies. A workload is a set of interdependent tasks that can be executed across different computing resources. The workflow manager can also apply specific optimization, such as parallelism for certain workflow parts. An example of multi-QPU parallelism is circuit knitting, which decomposes complex quantum circuits into smaller circuits~\cite{quantum_centric_hpc_2022}. Another example of task parallelism is ensemble parallelism, comprising loosely coupled tasks with minimal dependencies, e.\,g., found in parameter sweeps to evaluate a circuit with different parameters and when conducting parallel measurements.

\subsubsection{Workload Layer}

This workload layer orchestrates the execution of the tasks emitted by the workflow layer. The workload manager is the core entity of this layer and is responsible for selecting resources, partitioning the workload, and assigning tasks to resources~\cite{Turilli_2019}. We distinguish between classical and quantum workloads. A classical workload only comprises classical tasks, while quantum workloads are hybrid, containing both classical and quantum tasks.  Quantum workloads can be highly heterogeneous and hardware-dependent. For example, simulated, ion-trap, and superconducting QPUs have different runtime and fidelity trade-offs. 

Quantum-HPC systems face unique workload management and scheduling challenges, including (i) the lack of a unified standard for accessing QPUs and expressing hybrid workloads, (ii) the complex dependencies of applications to specific Quantum Processing Units (QPUs) that often necessitate manual, application-level adaptations, and (iii) the limited availability of physical QPUs that complicates the balancing between application-level and system-level objectives. Our conceptual middleware addresses these challenges by encapsulating the workload management and scheduling concerns while allowing for adequate information flows between application and Quantum-HPC systems.

The workload manager acquires the respective resources via the task layer. Considering the coupling between the quantum and classical tasks, it assigns and co-locates tasks to the resources. For example, classical tasks tightly coupled to quantum tasks, e.\,g., for the HPC-for-Quantum scenarios, must be co-allocated and assigned to resources nearby.

The scheduling of tasks requires both application- and system-level information~\cite{1392910}. Thus, application-level schedulers and Pilot-Jobs~
\cite{6404423,10.1145/3177851} may be crucial in bringing together application-level and system-level information. The assigning of tasks to resources is also referred to as binding. The binding of tasks can occur both early and late. Early binding directly assigns tasks to resources based on currently available information. Late binding allows for more dynamism and addresses, e.\,g, uncertainties like resource fluctuations and other variations in the infrastructure.

\subsubsection{Task Layer} 

The task layer is an integral component of the middleware system and comprises a collection of task managers. The task manager orchestrates the execution of tasks on a specific resource as assigned by the workload manager. A task manager is typically responsible for a single resource and manages the resource allocation, acquisition, scheduling, assignment, and monitoring to ensure the tasks run successfully. Typically, tasks are executed on HPC resources as part of a job or within a container on cloud resources. As described, a common mechanism to manage tasks across heterogeneous resources are Pilot-Jobs. %

The task manager also supports dynamic allocation during runtime, allowing for the acquisition and release of resources as needed. Additionally, the task manager is responsible for handling errors and failures that may arise during the execution of tasks. It ensures that the necessary resource requirements for successful task execution are met. In the context of quantum computing, resource co-location is crucial. For tightly coupled tasks, the QPU must be co-located with sufficient classical computing capacities, e.\,g., GPUs and CPU. 

For example, quantum error correction requires the co-location of QPU, GPU, and CPU due to the tight coupling between classical and quantum computing tasks. Variational algorithms also benefit from nearby classical computing resources, but the coupling does not occur within the coherence time of the QPU.

\subsubsection{Resource Layer} 

This layer represents the diverse HPC and cloud resources, such as classical computing (e.\,g., CPUs), QPUs, and accelerators (e.\,g., GPUs). While QPUs have been located remotely from the classical computation, increasingly tighter integrations of classical resources are emerging~\cite{RUEFENACHT2022,quantum_centric_hpc_2022}. For error correction, accelerated classical computing is required to perform the classical processing of the syndrome measurements. The increased complexity of the resource layer demands abstractions so that the resource layer is consistently presented. 

Parallelization of quantum workloads across nodes, cores, and accelerators (including QPUs) is critical to achieving the necessary performance and scale. For this purpose, the resource layer must integrate with the underlying HPC technologies, including QPU-specific compilers, GPU libraries (such as cuQuantum and cuTensor), and networks.

\subsection{Example: Quantum Chemistry Workflow}

In the following, we consider a quantum chemistry application where the task is to compute the ground state energy of a molecule, which is frequently used to predict the chemical properties of a molecule~\cite{doi:10.1021/acs.chemrev.8b00803}.

\paragraph{Workflow Layer} The Variational Quantum Eigensolver (VQE) is a NISQ algorithm for computing the ground state energy of a molecule. The ground state estimation is often part of an end-to-end workflow, which includes several pre- and post-computing steps  (\emph{Quantum-about-HPC} integration type). Examples of preparation steps are, e.\,g., reading the molecule data from a file and computing an approximate solution to the ground state using the Hartree-Fock (HF) method. 

The VQE algorithm itself is an example of the Quantum-in-HPC integration pattern. The quantum part is a parameterized quantum circuit, which executes on a QPU and estimates the ground state energy.
The classical component optimizes the parameters using a classical optimization algorithm. Further, enhancements emerged, e.\,g., embedding techniques that restrict the simulated quantum particles by utilizing classical simulation techniques. While this reduces the required qubits, it requires additional classical resources~\cite{10.1063/5.0029536}. Generally, this algorithm and the Quantum-in-HPC pattern require frequent communication between classical and quantum components as they iteratively update the parameters of the quantum circuit.

The application creates a high-level workflow description that interweaves quantum tasks (specifically a quantum circuit representing the molecular Hamiltonian) and classical tasks (optimization). This representation is the basis for efficiently managing dependencies (e.\,g., external libraries like Qiskit), inputs, execution, and outputs. The workflow description on this level is abstract, i.\,e., resource-independent. However, it is configurable to allow users some control (e.\,g., over resource types). 

Current quantum software frameworks (e.\,g., Qiskit and Pennylane) define workflows on a lower level involving precise implementation steps (e.\,g., in Python) and concrete resource mapping. 
While these frameworks provide some extension mechanisms, these are typically highly platform-specific and limited. For example, Qiskit's Provider API and Pennylane's Device API allow for integrating custom backends. Our conceptual middleware decouples the workflow description from the implementation details, and thus, enables optimization on the middleware level.

The middleware system is responsible for mapping the workflow description to a set of tasks and resources. For this purpose, the workflow manager resolves all dependencies and allocates the necessary resources. The output is a workload, i.\,e., the resources and tasks ready to run, which is forwarded to the workload manager.

\paragraph{Workload and Task Layer} 

The workload and task managers are essential to enable the scalable execution of quantum workflows and their associated workloads (objective \textbf{O-2}). The workload manager schedules and orchestrates the execution of quantum and classical tasks emitted by the workflow manager. For VQE, e.\,g., sufficient quantum and classical resources need to be allocated to optimize interactions between both components. The task manager is then responsible for the execution of individual tasks, either on the QPU or classical resources.

Depending on the characteristics of the workload, particularly the coupling between quantum and classical tasks, the tasks can be placed accordingly, ensuring performance and scale. For example, the pre- and post-computing steps of the workflow do not necessarily be co-located with the QPU. For current QPU capabilities, remote QPU access is sufficient for VQE. With the increasing scale, a co-allocation of resources is required to enable scalable variational algorithms. Further, a co-allocation is critical for HPC-for-Quantum scenarios, e.\,g., error mitigation routines.

\paragraph{Resource Layer} 

The emitted quantum and classical tasks are executed on the resource layer. To allow low-level resource-specific optimization (objective \textbf{O-3}), the execution typically involves just-in-time compilation steps to optimize the circuit for defined hardware. To evaluate the quantum state, repeated measurements are critical.

\section{Conclusion}
\label{sec:conclusion}

Managing and executing quantum workflows poses significant challenges, necessitating middleware systems to support applications in achieving scale.  To aid the design of middleware systems, we identified three integration patterns for quantum and HPC applications and characterized these, e.\,g., by application structure and coupling. Our analysis further shows the need for quantum HPC techniques (e.\,g., scalable and parallel execution, accelerators) arises in several parts, e.\,g., in the classical simulation of quantum circuits, variational algorithms, and quantum error correction. 

We propose a conceptual middleware system that enables seamless interaction between classical and quantum resources by providing unified resource access and management. The conceptual middleware decouples workload and task management from the application software while allowing low-level HPC optimization (e.\,g., resource-specific compilation steps, resource, and task co-allocation). It utilizes established high-performance computing abstractions and enables the seamless integration of quantum computing into HPC systems.

In future work, we aim to develop a reference implementation of the conceptual middleware. Further, we plan to explore the emulation of workloads and resources for hybrid quantum applications. The emulator development includes tools and techniques to mimic the characteristics of applications and infrastructure, thereby enabling the optimization of resource and task allocations before deploying them on actual quantum systems. Finally, the development of workflow and application-level benchmarks~\cite{9951297} will help to establish an understanding of the impact of quantum computing in end-to-end applications.

\section*{Acknowledgment}
We thank Pradeep Mantha for proofreading, valuable comments and discussions. The authors generated parts of this text with OpenAI's language-generation models. Upon generation, the authors reviewed, edited, and revised the language.

\ifCLASSOPTIONcaptionsoff
  \newpage
\fi

\bibliographystyle{IEEEtran}
\bibliography{literature}

\end{document}